# Closed-Form Solutions for Grid-Forming Converters: A Design-Oriented Study

Fangzhou Zhao, *Member, IEEE*, Tianhua Zhu, *Member, IEEE*, Lennart Harnefors, *Fellow, IEEE*,
Bo Fan, *Member, IEEE*, Heng Wu, *Member, IEEE*, Zichao Zhou, *Student Member, IEEE*,
Yin Sun, *Member, IEEE*, and Xiongfei Wang, *Fellow, IEEE*

*Abstract*— This paper derives closed-form solutions for grid-forming converters with power synchronization control (PSC) by subtly simplifying and factorizing the complex closed-loop models. The solutions can offer clear analytical insights into control-loop interactions, enabling guidelines for robust controller design. It is proved that 1) the proportional gains of PSC and alternating voltage control (AVC) can introduce negative resistance, which aggravates synchronous resonance (SR) of power control, 2) the integral gain of AVC is the cause of sub-synchronous resonance (SSR) in stiff-grid interconnections, albeit the proportional gain of AVC can help dampen the SSR, and 3) surprisingly, the current controller that dampens SR actually exacerbates SSR. Controller design guidelines are given based on analytical insights. The findings are verified by simulations and experimental results.

## I. INTRODUCTION

THE growing penetration of voltage-source converter (VSC)-based resources in electrical grids necessitates grid-forming (GFM) capabilities for VSCs [1], [2]. Differing from conventional grid-following VSCs, the GFM capabilities requires VSCs to behave as a voltage source behind an impedance and autonomously provide power responses to maintain the voltage and frequency of power grids [3]–[5].

There have been extensive studies on stability and control of GFM-VSCs. It is shown that when GFM-VSCs are connected to a highly inductive grid, the synchronous-frequency resonance (SR) may be manifested by the power synchronization control (PSC) [6]. To dampen the SR, a virtual resistance (VR) that is based on the feedback of VSC output current is used in [6], [7]. Designing the VR can be done by evaluating the damping ratio of closed-loop poles of PSC, which yields an empirical value of VR, i.e., 0.2 p.u. [8], yet this approach overlooks the dynamics of alternating voltage control (AVC). Recent studies in [9], [10] indicate that AVC can affect the PSC dynamics, which in turn can affect the damping effect of VR on SR. However, there is lack of analytical interpretation of how AVC impacts the SR, particularly when the current control (CC) is implemented.

In addition to the SR, the sub-synchronous resonance (SSR) issues are reported when GFM-VSC is connected to a stiff grid with a high short-circuit ratio (SCR) [11]–[16], where the SSR frequency is often around 10 Hz. Interactions caused by CC and AVC are respectively identified as the causes of SSR in [11] and [12], implying that the SSR phenomenon may vary with the design of inner control loops. The study in [11] presents a design-oriented analysis of SSR by investigating the poles of equivalent PSC plant in the closed-form. However, the impacts of AVC are still overlooked. By incorporating the AVC into a closed-loop model of GFM-VSC, the studies in [12]–[15] can fully characterize the SSR behavior, yet they are based on numerical studies, e.g., sensitivity analysis of control and circuit parameters. Consequently, the findings are case-specific, and the design guidelines of controllers tend to be empirical, which limits their generality for different operating conditions and system configurations.

This paper thus proposes a robust, analytical design approach for GFM-VSCs that help mitigate the risks of SR and SSR. The major contribution is the generalized closed-form solutions of the GFM control, considering different combinations of PSC, AVC and CC loops. Differing from numerical analysis, the analytical solutions not only shed insights into the conditions and causes of SR and SSR, but they also provide physical interpretations on impacts of different controllers, which enables to formulate a general guideline for GFM control design and parameter-tuning.

This paper is structured as follows. Section II describes the GFM-VSC system under study and the denotations of physical variables used in this work. Section III provides the closed-form solutions of GFM-VSC with PSC only. Section IV discusses closed-form solutions of GFM-VSC with both PSC and AVC, while Section V explores the control with PSC, AVC, and CC. A summary of the stability analysis of all control schemes is presented in Section VI, followed by experimental verifications in Section VII. Conclusions are finally drawn in Section VIII.

## II. SYSTEM DESCRIPTION

Fig. 1 shows a single-line diagram of the three-phase grid-connected VSC, and Table I presents the GFM control schemes for analysis. To reveal the impact of each control loop, which includes PSC, AVC, CC, and voltage feedforward (VFF), three schemes are considered in Table I to give a step-by-step study. The dc-link voltage is assumed to be constant in this study, as it can be regulated by a front-end converter or an energy storage system [11]. An $L$ filter is used with the VSC, and it is denoted by the inductance $L_f$ and equivalent series resistance (ESR) $R$. The grid is denoted by a voltage source behind an inductance $L_g$ [8], where $L_g$ is quantified by the SCR, seen from the point of common coupling (PCC), which is expressed as [17]

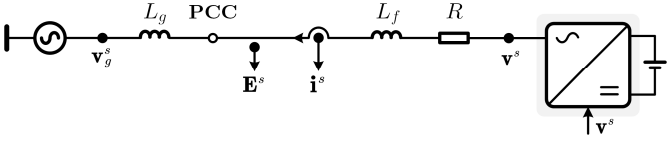

Fig. 1. Single-line diagram of a grid-connected three-phase VSC.

$$\text{SCR}=1/(\omega_1 L_g) \quad (1)$$

where $\omega_1$ is the fundamental frequency, and $\omega_1 L_g$ is calculated in per-unit (p.u.) value (base value will be introduced later).

The voltages and currents are denoted in stationary reference frame by complex space vectors [18], e.g., output voltage vector $\mathbf{E}^s$ and output current vector $\mathbf{i}^s$, and their corresponding denotations in the rotating $dq$-frame do not have the superscript $^s$ [9], e.g., $\mathbf{i}=i_d+ji_q$. In PSC, the active power $P$ is calculated by

$$P=\kappa\,\text{Re}\{\mathbf{E}^s(\mathbf{i}^s)^*\}=\kappa\,\text{Re}\{\mathbf{E}\mathbf{i}^*\},\quad \kappa=3/(2K^2) \quad (2)$$

where $\mathbf{E}^s$ and $\mathbf{E}$ are the PCC voltage vectors in the stationary frame and the $dq$-frame, respectively, and $K$ is the space vector scaling constant. For peak-value scaling ($K=1$), $\kappa=3/2$. For p.u. normalization of the quantities or power-invariant scaling ($K=\sqrt{3/2}$), $\kappa=1$ [8]. The superscript $^*$ in (2) denotes the complex conjugate.

In this work, all controls and modeling are conducted using p.u. values ($\kappa=1$), with the conclusions being consistent when translated into real values. The base value of all voltages is peak nominal phase voltage $E_B$. Likewise, the base current is peak nominal phase current $I_B$. Hence, the base impedance is given by $Z_B=E_B/I_B$, while base admittance is $Y_B=1/Z_B$. With (2), the base power is defined as $P_B=\kappa E_B I_B$.

The typical PSC controller – an integrator [6] – is used in this study, and $C_p(s)$ is defined as

$$C_p(s)=k_P/(\kappa s). \quad (3)$$

where the base value of the controller gain $k_P$ is $k_B=1/P_B$. $k_P$ is usually small, and a typical range is $0.01\omega_1 \sim 0.05\omega_1$ [1], [19].

The AVC in $dq$-frame often uses proportional integral (PI) controllers [20], and $C_v(s)$ is defined as

$$C_v(s)=G_a+k_i/s. \quad (4)$$

In scheme 2, the base value of $G_a$ and $k_i$ is 1, therefore the p.u. values and real values are the same. While in scheme 3, the base value of $G_a$ and $k_i$ is $Y_B=I_B/E_B$.

The CC and VFF are used to regulate fast current dynamics during faults [21]. The current limiter [22] between AVC and CC is not plotted for simplicity, as it does not affect small-signal dynamics. The CC in $dq$-frame also employs PI controllers. However, the I gain is typically small and is commonly omitted [10], [23]. Thereby, $C_i(s)$ is defined as

$$C_i(s)=R_a \quad (5)$$

where the base value of $R_a$ is $Z_B=E_B/I_B$.

In addition, the analysis is focused on the low-frequency (0~100 Hz) dynamics of GFM control. The cut-off frequencies of the low-pass filters (LPFs) used within AVC and VFF are often designed to be higher than 100 Hz [10]. Hence, they are not considered in Table I and hereafter.

### III. CLOSED-FORM SOLUTIONS FOR PSC

Scheme 1 is analyzed first, focusing only on PSC to reveal the characteristics without any inner-loop coupling. A constant voltage magnitude is used in Table I, and $V=1$ p.u.

The power stage model of Fig. 1 is given by

$$\mathbf{v}^s-(sL+R)\mathbf{i}^s=\mathbf{v}_g^s=V_g e^{j\omega_1 t} \quad (6)$$

where $\mathbf{v}^s$ and $\mathbf{v}_g^s$ are the converter and grid voltage vectors, respectively, and $L=L_f+L_g$ is the total inductance. Then, the

TABLE I
THREE SCHEMES OF GFM CONTROLS WITH DIFFERENT INNER LOOPS

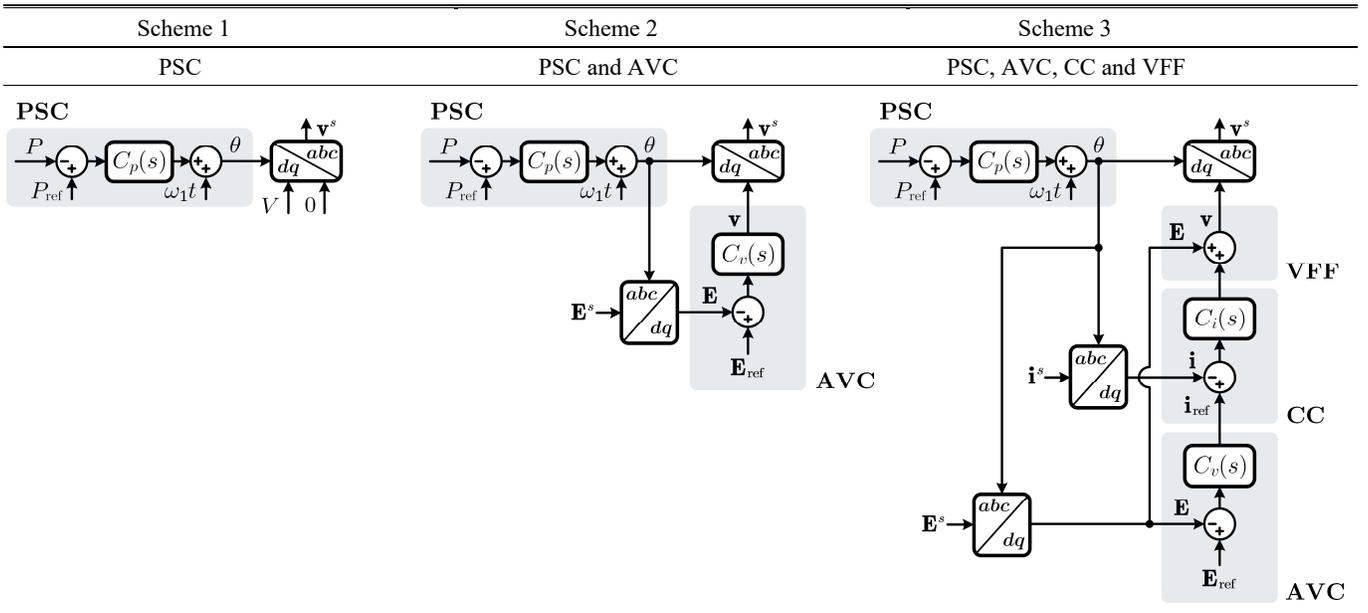

corresponding *dq*-frame vectors in (6) and rotating angle with small-signal perturbations are defined by

$$\theta = \omega_1 t + \theta_0 + \Delta\theta, \quad \mathbf{v} = \mathbf{v}_0 + \Delta\mathbf{v}, \quad \mathbf{i} = \mathbf{i}_0 + \Delta\mathbf{i} \quad (7)$$

where $\theta_0$, $\mathbf{v}_0 = V$ and $\mathbf{i}_0 = i_{d0} + ji_{q0}$ denote the static values. With (7), the small-signal model of (6) is derived as [8]

$$\Delta\mathbf{v} - [(s+j\omega_1)L + R]\Delta\mathbf{i} = j\{[(s+j\omega_1)L + R]\mathbf{i}_0 - V\}\Delta\theta \quad (8)$$

where the Laplace variable $s$ henceforth is to be considered as the operator $s = d/dt$, where appropriate. Since AVC and CC are not included, $\Delta\mathbf{v} = 0$. Thereby, (8) gives

$$\Delta\mathbf{i} = -j\frac{[(s+j\omega_1)L + R]\mathbf{i}_0 - V}{(s+j\omega_1)L + R}\Delta\theta. \quad (9)$$

By linearizing (2), the active power perturbation is given by

$$\Delta P = \kappa\,\mathrm{Re}\{\mathbf{E}_0\Delta\mathbf{i}^* + \mathbf{i}_0^*\Delta\mathbf{E}\}. \quad (10)$$

Similar to (8), the PCC voltage dynamics can be derived as

$$\Delta\mathbf{E} - (s+j\omega_1)L_g\Delta\mathbf{i} = j[(s+j\omega_1)L_g\mathbf{i}_0 - \mathbf{E}_0]\Delta\theta. \quad (11)$$

Substituting (9) and (11) into (10) gives the plant of PSC, i.e.,

$$\Delta P = \kappa\beta\frac{i_{q0}L_g s^2 + E_{q0}s + \omega_1\gamma}{(s+\alpha)^2 + \omega_1^2}\Delta\theta = G_{\theta P}(s)\Delta\theta \quad (12)$$

where $\alpha = R/L$, $\beta = V/L$ and $\gamma = E_{d0} + i_{q0}\omega_1 L_g$.

Based on (12), the closed-loop model of PSC is illustrated in Fig. 2, and the closed-loop transfer function is derived as

$$\Delta P = \frac{[k_P/(\kappa s)]G_{\theta P}(s)}{1 + [k_P/(\kappa s)]G_{\theta P}(s)}\Delta P_{\mathrm{ref}} = G_{\mathrm{PSC}}(s)\Delta P_{\mathrm{ref}},$$

$$G_{\mathrm{PSC}}(s) = \frac{k_P\beta(i_{q0}L_g s^2 + E_{q0}s + \omega_1\gamma)}{s^3 + (2\alpha + a_2)s^2 + (\omega_1^2 + \alpha^2 + a_1)s + k_P\beta\omega_1\gamma} \quad (13)$$

where $a_1 = k_P\beta E_{q0}$, $a_2 = k_P\beta i_{q0}L_g$, and $k_P$ is the integral gain.

Define the following coefficients $n_1 = 2\alpha$, $n_0 = \omega_1^2$, $m_2 = a_2$, $m_1 = \alpha^2 + a_1$ and $m_0 = k_P\beta\omega_1\gamma$, and substitute them into the denominator of $G_{\mathrm{PSC}}(s)$. Then, the three poles of $G_{\mathrm{PSC}}(s)$ can be solved by setting the denominator equal to 0, i.e.,

$$s^3 + (n_1 + m_2)s^2 + (n_0 + m_1)s + m_0 = 0 \quad (14)$$

Using the rules in Appendix I, the closed-form solutions of the three poles are approximately derived as

$$G_{\mathrm{PSC}}(s) = k_P\beta\frac{i_{q0}L_g s^2 + E_{q0}s + \omega_1\gamma}{(s - p_1)(s - p_2)(s - p_3)},$$

$$p_{1,2} \approx -\frac{2R - k_P V E_{d0}/\omega_1}{2L} \pm j\sqrt{\omega_1^2 - \left(\frac{2R - k_P V E_{d0}/\omega_1}{2L}\right)^2}$$

$$\approx -\frac{2R - k_P V E_{d0}/\omega_1}{2L} \pm j\omega_1,$$

$$p_3 \approx -k_P(E_{d0} + i_{q0}\omega_1 L_g)V/(\omega_1 L). \quad (15)$$

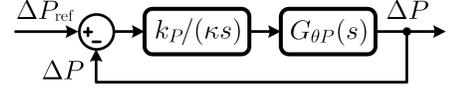

Fig. 2. Small-signal model of PSC loop.

TABLE II
PARAMETERS OF POWER CIRCUIT

| Symbol | Description | Value |
| --- | --- | --- |
| $E_n$ | Rated voltage (L-L, RMS) | 190.5 V (1 p.u.) |
| $P_n$ | Rated capacity | 5 kW (1 p.u.) |
| $\omega_1$ | Nominal frequency | 314 rad/s (50 Hz) |
| $L_f$ | Filter inductance | 3 mH (0.1298 p.u.) |
| $R$ | ESR of filter | 189 m$\Omega$ (0.026 p.u.) |

The details are shown in Appendix I, which offers a method to simplify the cubic equation for factorization. The coefficients in this scheme can satisfy the required conditions (35) and (37) in Appendix I to obtain the approximated solutions. The model (15) is verified as follows.

*Example 1:* Consider a power circuit in Fig. 1 and parameters listed in Table II. The VSC is controlled in scheme 1 in Table I and $k_P = 0.03\omega_1$.

Fig. 3 compares the frequency responses of proposed model (15) against the frequency scan results in EMT simulations for SCR=2 and SCR=10, respectively. The frequency scan results agree well with the model, which verifies the derivations.

In (15), the damping ratio of $p_{1,2}$ can be further derived by

$$\zeta_{1,2} \approx \frac{\omega_1 R - k_P V E_{d0}/2}{L}. \quad (16)$$

***Remark 1:*** The following observations can be made based on (15) and (16):
1) The resistance $R$ is important to guarantee the stability. For a VSC system with a high $X/R$ ratio, $\zeta_{1,2}$ are small and $p_{1,2}$ tend to be under-damped.
2) The PSC gain $k_P$ negatively affects $\zeta_{1,2}$, hence a high $k_P$ can destabilize the system. The maximum value $k_{P\max}$ that is critical to the system stability can be derived as

$$k_{P\max} = 2R\omega_1/(VE_{d0}). \quad (17)$$

Given that both $V$ and $E_{d0}$ are near 1 p.u., it can be inferred that $k_{P\max}$ is around $2R\omega_1$ in p.u. value (a typical droop selection is $0.03\omega_1$). The grid resistance, if considered, can be regarded as a part of $R$ in (17). Thus, $k_{P\max}$ becomes higher with the increase of grid resistance. In this study, Fig. 1 only considers the ESR of $L$-filter, which gives the worst-case design for $k_{P\max}$.
3) In the imaginary parts of $p_{1,2}$ in (15), $\omega_1^2$ is much larger than the other term. Thus, the oscillation frequency of $p_{1,2}$ is around the synchronous frequency $\omega_1$, i.e. the SR mode.
4) Provided that $k_P < k_{P\max}$, the system can remain stable under both weak- and stiff-grid conditions even if the damping is insufficient. Different grid inductance and filter $L = L_g + L_f$ in the denominator of (16) will not shift the poles to the right-half plane (RHP).

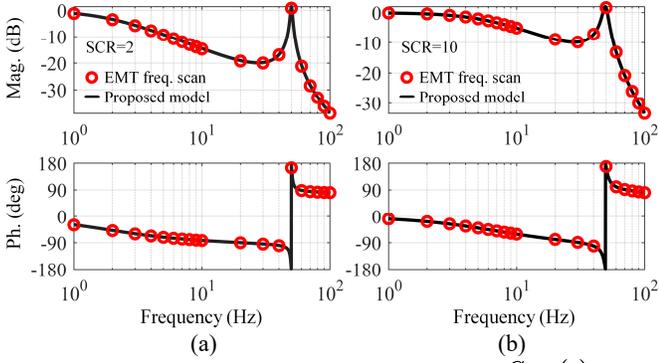

Fig. 3. Comparison of closed-loop transfer function $G_{\text{PSC}}(s)$ in (15) and EMT simulation model. (a) SCR=2. (b) SCR=10.

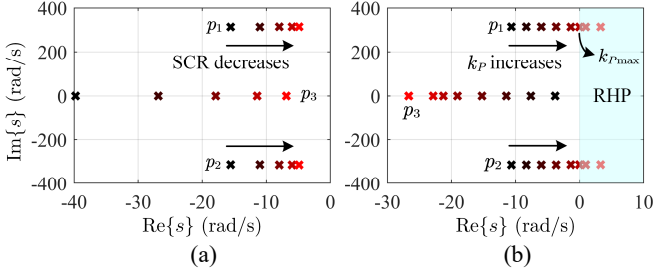

Fig. 4. Pole-plot of $G_{\text{PSC}}(s)$. (a) SCR decreases from 10 to 1.5. (b) $k_P$ increases from $0.01\omega_1$ to $0.07\omega_1$.

Following *Example 1*, Fig. 4 gives the pole-plots of $G_{\text{PSC}}(s)$. The conjugate poles $p_{1,2}$ move to the right as SCR decreases in Fig. 4(a). However, with $k_P=0.03\omega_1 < k_{P\max}=0.0558\omega_1$, $p_{1,2}$ will not enter the RHP, causing instability. Therefore, the system is stable irrespective of SCR. In Fig. 4(b), the increase of $k_P$ (from $0.01\omega_1$ to $0.07\omega_1$) degrades the damping ratio. Yet, when $k_P > k_{P\max}$, $p_{1,2}$ enter the RHP, the system is unstable.

## IV. CLOSED-FORM SOLUTIONS FOR PSC AND AVC

This section presents closed-form solutions for GFM control with PSC and AVC (scheme 2 in Table I). The AVC uses a PI controller, i.e., $C_v(s)=G_a+k_i/s$. As will be proved as follows, the P gain $G_a$ shows negative damping on SR, which is the main risk of instability. The I gain $k_i$ is selected from [24], giving $k_i \approx \omega_c/10$, where $\omega_c$ is the crossover frequency. To make a fair comparison with the final scheme 3, where the bandwidth of AVC is lower than inner loop CC, e.g., $10\omega_1$ rad/s [25], it is thus assumed that $\omega_c \leqslant 10\omega_1$. Then, $k_i$ is set lower than $\omega_1$ in p.u. value, and it shows limited effect on SR compared to $G_a$. Hence, I controllers are omitted in this model for simplicity, but we will demonstrate its influence in experiments in Section VII.

Then, the dynamics of control voltage with AVC is given by

$$\Delta \mathbf{v} = -G_a \Delta \mathbf{E}. \tag{18}$$

(18) is different from scheme 1 where the open-loop voltage control is used ($\Delta \mathbf{v}=0$), but the rest of the modeling process to obtain the closed-form solutions is similar. For clarity, the complete derivations are given in Appendix II, and the results are presented as follows. It is found that the AVC can equivalently reshape the plant in Fig. 2, and therefore modify the closed-loop transfer function of PSC $G_{\text{PSC}a}(s)$ as

$$G_{\text{PSC}a}(s) = \frac{N_a(s)}{(s-p_{1a})(s-p_{2a})(s-p_{3a})} \tag{19}$$

where $N_a(s)$ is the nominator (detailed expression is given in Appendix II), and it also has three poles ($p_{1a}$, $p_{2a}$ and $p_{3a}$). From Appendix II, the closed-form solutions of these poles can be approximated to

$$p_{1,2a} \approx -\frac{R-\delta}{L+G_a L_g} \pm j\omega_1, \quad \delta = \frac{k_P(VE_{d0}+G_a|\mathbf{E}_0|^2)}{2\omega_1},$$

$$p_{3a} = -\frac{2\delta + k_P L_g(Vi_{q0}+G_a \text{Im}\{\mathbf{E}_0^* \mathbf{i}_0\})}{L+G_a L_g}. \tag{20}$$

The model is verified by the following example.

*Example 2:* Consider a power circuit in Fig. 1 and parameters listed in Table II. The VSC is controlled in scheme 2 in Table I. The control parameters are $k_P=0.03\omega_1$ and $C_v(s)=G_a=0.5$ p.u. Fig. 5 compares the model $G_{\text{PSC}a}(s)$ with poles calculated in (20) and the frequency scan results in EMT simulations when SCR=2 and SCR=10, respectively. The proposed model (20) agrees well with the simulation results.

Based on (20), the damping ratios of $p_{1,2a}$ can be given by

$$\zeta_{1,2a} \approx \frac{R-\delta}{\omega_1(L+G_a L_g)} \tag{21}$$

where $\delta = k_P(VE_{d0}+G_a|\mathbf{E}_0|^2)/(2\omega_1)$ is given in (20).

*Remark 2:* The following observations can be made based on (20) and (21):
1) The P gain $G_a$ of AVC has two effects on $\zeta_{1,2a}$. First, $G_a$ contributes to more negative resistance $\delta$ in the numerator, similar to PSC gain $k_P$. Second, it adds a virtual inductance in the denominator. Both effects decrease the damping ratio, and the risk of SR evidently increases compared to (16), where AVC is not used.
2) Due to the negative effects of AVC, the maximum control gain of PSC $k_P$ is decreased. To guarantee the stability, i.e., $R-\delta>0$, $k_{P\max}$ is obtained by

$$k_{P\max} = 2R\omega_1/(VE_{d0}+G_a|\mathbf{E}_0|^2). \tag{22}$$

By assuming $VE_{d0} \approx |\mathbf{E}_0|^2 \approx 1$ p.u., an estimation of $k_{P\max}$ is $k_{P\max}=2R\omega_1/(1+G_a)$. As a result, the P controllers of AVC aggravate SR, and thus tend to destabilize the system.

Following *Example 2*, Fig. 6 shows the pole-plots of the model $G_{\text{PSC}a}(s)$ when $G_a$ and $k_P$ increase, respectively, where $p_{1,2a}$ move to the right and finally enter RHP in both cases. Fig. 8(a) shows that $G_a$ has negative damping on SR. In Fig. 6(b), $k_{P\max}=0.0362\omega_1$ is lower than that of *Example 1*, where AVC is not used in Fig. 4(b) ($k_{P\max}=0.0558\omega_1$).

## V. CLOSED-FORM SOLUTIONS FOR PSC, AVC, CC AND VFF

This section presents closed-form solutions of GFM control, including PSC, AVC, CC and VFF (scheme 3 in Table I). To highlight the effects of inner loop controllers on different resonances, we will discuss the AVC with P controllers first, and then come to the PI controllers for a step-by-step analysis.

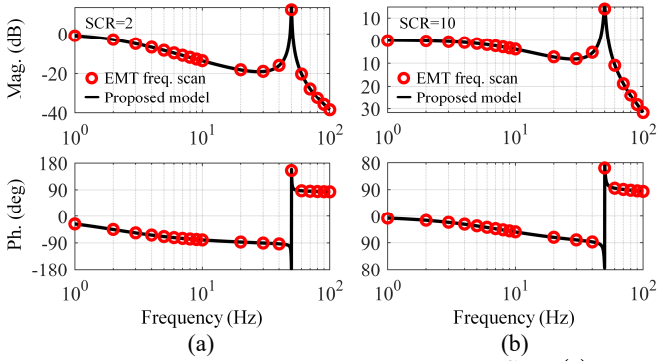

Fig. 5. Comparison of closed-loop transfer function $G_{\text{PSC}a}(s)$ in (19) with poles (20) and EMT simulation model. (a) SCR=2. (b) SCR=10.

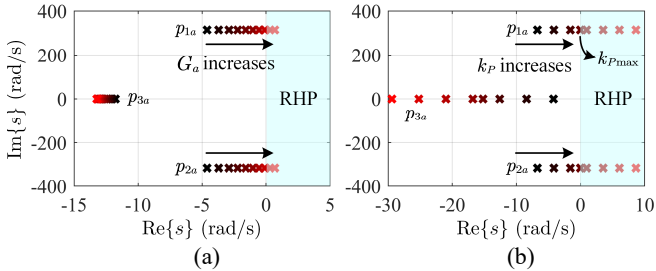

Fig. 6. Pole-plot of $G_{\text{PSC}a}(s)$. (a) $G_a$ increases from 0.1 to 1.0 p.u. (b) $k_P$ increases from $0.01\omega_1$ to $0.07\omega_1$.

### A. AVC with P controller

In this scheme, the P gains of AVC and CC are set as $G_a$ and $R_a$, respectively. The bandwidth of CC loop typically ranges from hundred Hz to a few kHz, which results in a high P gain $R_a$ (around 1 p.u., which is much higher than $R$). For instance, when $L_f = 0.1$ p.u., $R_a$ is set to 1 p.u. to achieve a bandwidth of 500 Hz for CC [25].

Then, the dynamics of control voltage $\Delta\mathbf{v}$ with AVC, CC and VFF are expressed as

$$\Delta\mathbf{v} = R_a(\Delta\mathbf{i}_{\text{ref}} - \Delta\mathbf{i}) + \Delta\mathbf{E} = (-R_aG_a+1)\Delta\mathbf{E} - R_a\Delta\mathbf{i}. \quad (23)$$

Appendix III gives the rest of the modeling process, which is similar to the previous Sections. The final closed-loop transfer function $G_{\text{PSC}b}(s)$ that incorporates the dynamics of AVC, CC and VFF can be obtained as

$$G_{\text{PSC}b}(s) = \frac{N_b(s)}{(s-p_{1b})(s-p_{2b})(s-p_{3b})} \quad (24)$$

where $N_b(s)$ is the nominator (detailed expression is given in Appendix III), and it also has three poles ($p_{1b}$, $p_{2b}$ and $p_{3b}$). From Appendix III, the closed-form solutions of the conjugate poles can be approximated to

$$p_{1,2b} \approx -\frac{R+R_a}{L+(R_aG_a-1)L_g} \pm j\omega_1 \quad (25)$$

The solution of real pole $p_{3b}$ is given in (49).

The model is verified by the following example.

*Example 4:* Consider a power circuit in Fig. 1 and parameters listed in Table II. The VSC is controlled in scheme 3 in Table I. The control parameters are $k_P = 0.03\omega_1$, $R_a = 0.865$ p.u. and $G_a = 3$ p.u.

Fig. 7 compares the proposed model, i.e., $G_{\text{PSC}b}(s)$ with derived poles $p_{1,2b}$ in (25), and the frequency scan results in EMT simulations when SCR=2 and SCR=10, respectively. The frequency scans can validate the closed-form solutions.

*Remark 3:* The following observations can be made based on (25):

1) The CC adds a virtual resistance $R_a$ to the real part of $p_{1,2b}$, which effectively dampens SR.
2) The AVC gain $G_a$ equivalently adds a virtual inductance to the real parts of $p_{1,2b}$, and decreases the damping ratio for SR. However, the VFF, which yields the term "-1" in the denominator of (25), has an opposite effect.
3) With the decrease of SCR, the damping ratio $\zeta_{1,2c}$ of $p_{1,2b}$ decreases, and the worst case is $L_g = 1$ p.u. To guarantee that $\zeta_{1,2c}$ is higher than 0.707, $R_a$ and $G_a$ can be designed based on (25) following

$$(1-G_a/\text{SCR})R_a \gtrsim \omega_1 L_f \quad (26)$$

where (26) is in p.u. values and SCR is given in (1).

4) With a proper design of $R_a$ and $G_a$ in 3), the control shows robust stability irrespective of SCR, especially the stiff grid, where the resonance of $p_{1,2b}$ is fully damped.

Following *Example 4*, Fig. 8 shows the poles of $G_{\text{PSC}b}(s)$. In (a), $G_a = 3$ p.u. and $R_a$ increases from 0.1 to 1.0 p.u. The CC can offer evident damping to $p_{1,2b}$ and mitigate the risk of SR. This allows a higher AVC gain $G_a$ as shown in (b), compared to the case without CC in Fig. 6(a), where the system becomes unstable when $G_a$ increases to 1 p.u.

### B. AVC with PI controller

To minimize the steady-state error of PCC voltage magnitude, the AVC needs PI controllers $C_v(s) = G_a + k_i/s$. The I gain $k_i$ is often set lower than $\omega_1$ in p.u. value, as explained in Section IV, and it shows limited effects on SR. Therefore, when CC ($R_a$) is used, the SR can be effectively damped.

However, as will be proved in the following analysis, the AVC I controllers can lead to a new risk of SSR. In this scheme, it is more difficult to solve the full-order model for any given operating conditions.

Therefore, to identify the SSR risk by closed-form solutions, we focus more on the light-load operating conditions where $\mathbf{i}_0 \approx 0$, and the key influencing factors of SSR will be identified theoretically – this is the main contribution of this subsection. For other load conditions, the SSR risk may still persist, but the closed-form solutions are too complicated to shed any insights, which deserves more studies in future.

In this case, the dynamics of $\Delta\mathbf{v}$ with AVC, CC and VFF are given as

$$\begin{aligned}\Delta\mathbf{v} &= R_a(\Delta\mathbf{i}_{\text{ref}} - \Delta\mathbf{i}) + \Delta\mathbf{E} \\ &= [-R_a(G_a+k_i/s)+1]\Delta\mathbf{E} - R_a\Delta\mathbf{i}.\end{aligned} \quad (27)$$

Appendix IV gives the detailed models of converter under light load conditions ($\mathbf{i}_0 \approx 0$). Additionally, since SR is well-damped by CC, we focus more on SSR in the lower-frequency range (e.g., $f<25$ Hz) to simplify the model. The final closed-loop transfer function $G_{\text{PSC}c}(s)$ that incorporates the dynamics of AVC, CC and VFF can be obtained as

$$G_{\text{PSC}c}(s)|_{f<25\text{ Hz}} \approx \frac{N_c(s)}{(s-p_{1c})(s-p_{2c})(s-p_{3c})} \quad (28)$$

where $N_c(s)$ is the nominator (detailed expression is given in Appendix IV), and it also has three poles ($p_{1c}$, $p_{2c}$ and $p_{3c}$). In (28), the poles of SR (already dampened by CC) are omitted in this frequency range ($f<25$ Hz). From Appendix IV, the closed-form solutions of $p_{1c}$, $p_{2c}$ and $p_{3c}$ can be approximated to

$$p_{1,2c} = (\rho_1-\rho_2)\frac{R_a k_i L_g}{2L_{eqc}} - \frac{k_P G_a \psi}{2}$$
$$\pm j\sqrt{\frac{k_i k_P \psi}{\rho_1} - \left[(\rho_1-\rho_2)\frac{R_a k_i L_g}{2L_{eqc}} - \frac{k_P G_a \psi}{2}\right]^2}, \quad (29)$$

$$p_{3c} = -\frac{R_a k_i L_g}{L_f + 2R_a L_g G_a}$$

where

$$L_{eqc} = L_f + R_a G_a L_g, \quad \psi = \frac{V^2 R_a/\omega_1 L_{eqc}}{\left[\frac{R_a(1+k_i L_g)}{\omega_1 L_{eqc}}\right]^2 + 1},$$

$$\rho_1 = \frac{L_f + R_a L_g G_a}{L_f + 2R_a L_g G_a}, \quad \rho_2 = \frac{2}{\left[\frac{R_a(1+k_i L_g)}{\omega_1 L_{eqc}}\right]^2 + 1}, \quad (30)$$

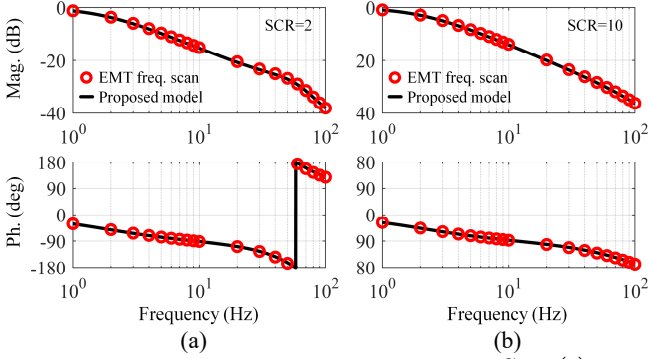

Fig. 7. Comparison of closed-loop transfer function $G_{\text{PSC}b}(s)$ in (24) with poles (25) and EMT simulation model. (a) SCR=2. (b) SCR=10.

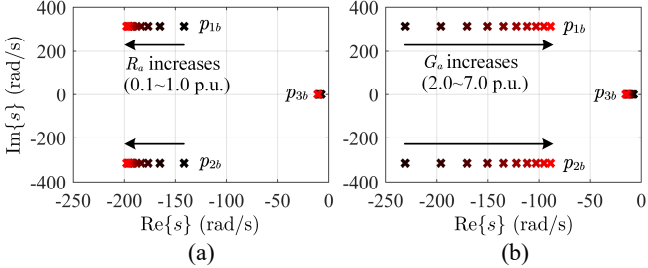

Fig. 8. Pole-plots of $G_{\text{PSC}b}(s)$. (a) CC gain $R_a$ increases from 0.1 to 1.0 p.u. (b) AVC gain $G_a$ increases from 2.0 to 7.0 p.u.

The model is verified by the following example.

*Example 5:* Consider a power circuit in Fig. 1 and parameters listed in Table II. The VSC is controlled in scheme 3 in Table I. The control parameters are $k_P=0.03\omega_1$, $R_a=0.865$ p.u., $G_a=3$ p.u. and $k_i=100$ p.u.

Fig. 9 compares the proposed model, i.e., (28) and (29), and EMT simulation results when SCR=2 and SCR=10, respectively. The proposed model with proper simplifications shows good accuracy within the focused frequency range below 25 Hz. Therefore, it is sufficient to characterize the SSR issues such as the resonance peak under strong grids conditions in Fig. 9(b).

*Remark 4:* The following observations can be made based on (29) and (30):

1) When the grid is stiff, the oscillation frequency of $p_{1,2c}$ (imaginary part) satisfies

$$\text{Im}\{p_{1,2}\} = \sqrt{\frac{k_i k_P \psi}{\rho_1} - \left[(\rho_1-\rho_2)\frac{R_a k_i L_g}{2L_{eqc}} - \frac{k_P G_a \psi}{2}\right]^2}$$
$$< \sqrt{\frac{k_i k_P \psi}{\rho_1}} < \sqrt{0.25 k_i \omega_1 V^2}. \quad (31)$$

See Appendix V for complete derivations. To obtain (31), the parameters are assumed to be reasonably designed, e.g., a low droop value $k_P \leqslant 0.05\omega_1$ [1] and a proper CC gain $R_a \approx 1$ p.u. to give a fast CC bandwidth around 500 Hz [23]. In practice, a low value of $k_i$ is often used. If we assume $k_i < \omega_1$ and $V \approx 1$ p.u., the oscillation frequency satisfies

$$\omega_{1,2c} < 0.5\omega_1 \quad (32)$$

(32) indicates the resonance is SSR (< 25 Hz).

2) Since $\rho_1$, $\rho_2$ and $\psi>0$, the system will be unstable ($p_{1,2c}$ move to RHP) only when $\rho_1-\rho_2>0$.
   As the grid impedance $L_g$ decreases, $\rho_1$ tends to be 1, while $\rho_2$ decreases and is much closer to 0. This gives a high risk of instability (positive real parts of $p_{1,2c}$). The stiff-grid conditions can cause the SSR introduced by $p_{1,2c}$.
   Additionally, increasing CC gain $R_a$ can further reduce $\rho_2$ that may cause SSR. This behavior is opposite to the damping of SR, as proved in Section V-A.

3) Increasing AVC P gain $G_a$ is helpful to move $p_{1,2c}$ to the left and thus mitigate the SSR, as indicated by the real part in (29). This behavior is also opposite to the damping of SR, as proved in Section V-A.

4) The system is always stable when $\rho_1-\rho_2\leqslant 0$. Considering that $\rho_1<1$, one can find a sufficient condition to estimate stability, i.e., $\rho_2>1$. Then substituting (30) into $\rho_2>1$, an interesting conclusion is given as follows – by selecting the parameters $G_a-k_i/\omega_1>0$, the system is guaranteed to be stable when SCR meets

$$\text{SCR} < \frac{G_a-k_i/\omega_1}{1-\omega_1 L_f/R_a}. \quad (33)$$

See Appendix VI for proof. (33) is in p.u. value, and it is a sufficient condition. If $R_a \gg \omega_1 L_f$, then $\omega_1 L_f/R_a$ in (33) can also be omitted. As a result, a design guideline is to set $G_a-k_i/\omega_1>5$, which guarantees stability for any SCR below 5. This is certainly a conservative design.

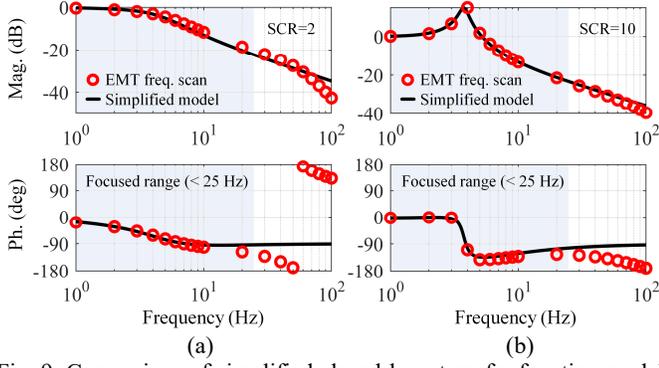

Fig. 9. Comparison of simplified closed-loop transfer function model in (28) with poles (29) and EMT simulation model within a focused frequency range < 25 Hz. (a) SCR=2. (b) SCR=10.

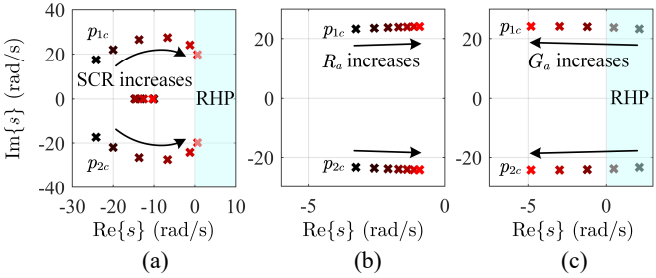

Fig. 10. Plots of poles $p_{1,2c}$. (a) SCR increases from 1.5 to 20. (b) $R_a$ increases from 0.4 to 1.0 p.u. (c) $G_a$ increases from 2 to 4 p.u.

Following *example 5*, Fig. 10 shows the poles of $G_{\text{PSC}c}(s)$ with respect to SCR, $R_a$ and $G_a$, respectively. When SCR increases, $p_{1,2c}$ move to the right and finally enter RHP in Fig. 10(a), indicating instability of stiff-grid connections. Increasing CC gain $R_a$ degrades the damping ratio in Fig. 10(b). However, a higher AVC P gain $G_a$ can shift the poles to the left, as shown in Fig. 10(c), which effectively enhances the damping ratio. The effects of $R_a$ and $G_a$ on SSR in Fig. 10(b) and (c) are exactly opposite to Fig. 8(a) and (b) on SR. These characteristics agree well with the insights above.

## VI. SUMMARY AND COMPARISON

Table III presents a summary of the instability risks of GFM controls with respect to the 3 schemes analyzed above. There are two instability risks – the SR (oscillation frequency around 50 Hz) and the SSR (oscillation frequency lower than 25 Hz).

On the one hand, the occurrence of SR is often attributed to an insufficient level of system resistance (a high $X/R$ ratio), which limits the PSC gain $k_P$ in scheme 1. The P controllers ($G_a$) of AVC in scheme 2 further decrease the damping to SR and leads to a higher instability risk. In scheme 3, the P controllers ($R_a$) of CC act as equivalent resistance, which can effectively dampen SR. Hence, under the condition that $R_a$ is adequate, scheme 3 is devoid of the SR risk.

On the other hand, scheme 3 shows the SSR risk under light-load conditions, which is caused by the interactions between PSC and AVC integrators in stiff-grid connections. In this case, increasing $G_a$ or decreasing $R_a$ helps to mitigate the resonance. However, it is suggested to consider both SR and SSR when choosing these parameters since they have opposing effects. For instance, an excessive value of $R_a$ to dampen SR may, in turn, exacerbates SSR.

The design guidelines for key controller parameters are also summarized in Table III based on the closed-form models. Note that the conditions (17) and (22) for scheme 1 and 2 are both necessary and sufficient. While the condition (33) for scheme 3 is only sufficient and gives a conservative design. In practice, it is not possible to use an excessive value of $G_a$ to stabilize the system with very high SCRs. To break through the limitations of stiff-grid connections, more enhanced control methods can be employed, e.g., [10], [26], [27].

## VII. EXPERIMENTAL VERIFICATION

Fig. 11 shows the diagram of experimental setup, and the parameters of power circuit are listed in Table II. The controller is dSPACE DS1007. An $LC$-filter is employed ($C$=10 μF), and the dc-link voltage is maintained by an independent dc source. The 3 control schemes in Table I are evaluated respectively under both stiff and weak grid conditions.

TABLE III
SUMMARY OF INSTABILITY RISKS AND DESIGN GUIDELINES OF GFM CONTROLS WITH RESPECT TO 3 SCHEMES

| Scheme (different controls) | Instability risk (SR or SSR), condition and cause | Design guidelines for controller parameters |
| --- | --- | --- |
| Scheme 1<br>PSC ($k_P/s$) | **SR (50 Hz).**<br>Condition – system with a high $X/R$ ratio.<br>Cause – Insufficient resistance in PSC plant and virtual negative resistance caused by $k_P$. | Select $k_P < k_{P\max} = 2R\omega_1/(VE_{d0})$ in (17).<br>Damping to SR can be enhanced by decreasing $k_P$. |
| Scheme 2<br>PSC ($k_P/s$) and AVC ($G_a+k_i/s$) | **SR (worse than scheme 1).**<br>Condition – system with high $X/R$ ratio.<br>Cause – Insufficient resistance in PSC plant and virtual negative resistance caused by both $k_P$ and $G_a$. | Select $k_P$ and $G_a$ to meet<br>$k_P < k_{P\max} = 2\omega_1 R/(VE_{d0}+G_a|\mathbf{E}_0|^2)$ in (22).<br>Damping to SR can be enhanced by decreasing $k_P$ or $G_a$. |
| Scheme 3<br>PSC ($k_P/s$), AVC ($G_a+k_i/s$) and CC ($R_a$) | **SSR (CC can effectively dampen SR).**<br>Condition – strong grids.<br>Cause – interactions between PSC and AVC integrators, which introduce new poles and resonance (< 25 Hz). | Select $G_a$ and $k_i$ to meet $G_a - k_i/\omega_1 > 5$ in (33) to ensure stability when SCR<5 (a conservative design).<br>Damping to SSR can be enhanced by increasing $G_a$, or decreasing $R_a$ or $k_i$. |

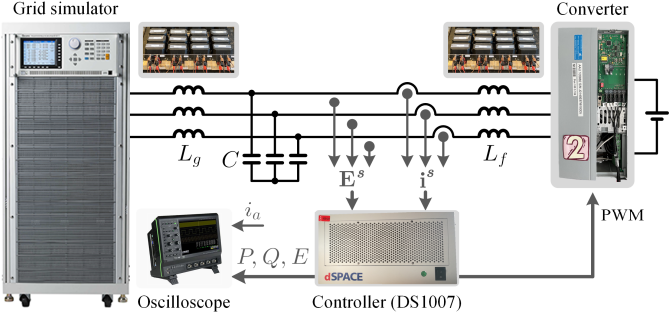

Fig. 11. Diagram of experimental setup.

### A. Stiff grid (SCR=11.6)

Fig. 12 presents a detailed comparison of the active power step responses for the aforementioned controls when SCR=11.6 ($L_g$ = 2 mH).

In Fig. 12(a), the PSC in scheme 1 reveals clear active power oscillation around 50 Hz, as shown in the expanded view ($\approx 20$ ms). The observed SR can be tracked back to the under-damped conjugate poles $p_{1,2}$ identified in Section III.

The performance of PSC and AVC ($G_a$= 0.15 p.u.) is shown in Fig. 12(b). Despite the use of a low value of $G_a$, the damping to power oscillation decreases evidently when compared to PSC in Fig. 12(a). The P controllers ($G_a$) of AVC results in a degradation of the damping ratio. The explanation for this phenomenon can be found in Section IV-A.

In Fig. 12(c) and (d), we further evaluate the influence of AVC integrators ($k_i/s$) and PSC. The different AVC integral gains – $k_i$= 25 p.u. and $k_i$= 65 p.u. – are compared. Both of them present clear power oscillations around 50 Hz. By comparing (c) and (d), it is seen that increasing $k_i$ can cause a deterioration in damping. As a result, both P and I controllers in AVC aggravate the SR. Typically, $k_i$ can be kept low to avoid the influence on SR, whereas $G_a$ shows more obvious impacts near 50 Hz.

Fig. 12(e) shows the responses of scheme 3 when AVC only uses P controllers. In this case, $G_a$= 2 p.u. and a low value of $R_a$= 0.2 p.u. is chosen to avoid a significant static error of AVC due to the absence of AVC integrators. By comparing (a), (b) and (e), it can be found that the SR is effectively suppressed by the CC, even when the value of $G_a$ in (e) is much higher than in case (b). This provides further evidence to support the effectiveness of the CC in offering equivalent resistance and critical damping to SR, as proved in Section V-A. Note that this case is mainly used to demonstrate the effects of CC, whereas in practice, AVC often employs both P and I controllers.

Fig. 12(f) shows the response of scheme 3 when AVC uses PI controllers. The bandwidth of CC is set to 6.7 p.u., giving $R_a$= 0.865 p.u. according to [23]. The parameters of AVC PI controllers are $G_a$= 3 p.u. and $k_i$= 100 p.u. The SR is well-dampened by CC, however, an even more serious SSR around 9 Hz (110 ms) is observed. This finding is consistent with the conclusions in Section V-B – the new resonance is introduced by $p_{1,2c}$ under stiff-grid conditions, which occurs below 25 Hz and originates from the I controllers of AVC. The emergence of $p_{1,2c}$ poses a potential risk of SSR, especially when SCR varies over a wide range. The CC can enhance the damping to SR, but not SSR.

Moreover, to validate the guidelines on parameter tuning of scheme 3, Fig. 13 presents a comparison of their active power responses. As Fig. 13(a) shows, decreasing $R_a$ from 0.865 to 0.346 p.u. can mitigate the low-frequency oscillation compared to Fig. 12(f). A low CC bandwidth is desirable from this point. However, it is a common practice to limit the overcurrent in a fault event by employing a fast CC, which often gives high gain of $R_a$ near 1 p.u. This trade-off should be considered. Another way to dampen SSR is to increase $G_a$, as shown in Fig. 13(b). Compared to Fig. 12(f), when $G_a$ increases from 3 to 5 p.u. and $R_a$ still equals to 0.865 p.u., the oscillation evidently reduces. However, compared to Fig. 13(b), when $k_i$ rises from 100 to 350 p.u. and maintaining $G_a$=5 p.u., the damping ratio to SSR becomes insufficient again. Hence, it is beneficial to select a higher AVC P gain ($G_a$) but keep a relatively low value of I gain ($k_i$). These results can corroborate the remarks in Section V-B.

### B. Weak grid (SCR=1)

Fig. 14 shows the comparison of active power responses of all schemes in Table I when SCR=1 ($L_g$=23 mH). The $X/R$ ratio of grid impedance is around 5, and this weak grid provides much more resistance to SR than the stiff grid does (almost 10 times than the stiff-grid case). As a result, PSC in Fig. 14(a) presents no oscillation near fundamental frequency, unlike in Fig. 12(a) in stiff grids. The grid resistance can mitigate the negative damping caused by $k_P$ in PSC, as indicated by (16). Fig. 14(b) shows the results of scheme 2 when $G_a$=0.15 p.u. and $k_i$= 25 p.u. In virtue of the grid resistance, there is no SR issue either, compared to Fig. 12(b) and (c). Further, in scheme 3, Fig. 14(c) demonstrates that the risk of SSR is much lower under weak grids conditions than stiff grids as shown in Fig. 12(f), which can also verify the analysis of SSR in Section V-B. The conjugate poles have adequate damping in all three cases when the SCR is low, and the real pole determines the power dynamic performances. Consequently, when the $X/R$ ratio is not sufficiently high (e.g., $X/R < 20$), the stiff-grid connection of GFM converter is more prone to resonances (SR or SSR) than weak ones.

## VIII. CONCLUSION

This paper fully investigates the dynamics of PSC coupled with AVC and CC by deriving closed-form solutions of poles to gain insights into designing stable GFM controls. To sum up, the primary results can be outlined as follows.

1) Resistance insufficiency is the cause of SR. The PSC gain and AVC P controller gain introduce negative resistance to SR, which degrades the damping.
2) CC can provide critical equivalent resistance to counteract the effects above and dampen SR.
3) The SSR is essentially caused by AVC integrators under high SCR conditions (e.g., SCR>10).
4) CC presents negative damping to SSR, though it mitigate SR. P gain of AVC can alleviate SSR, though it exacerbates SR. There are thus trade-offs. The bandwidth of CC is often high to regulate overcurrent during faults. Hence, to reduce the risk of SSR, it is suggested to properly increase the P gain of AVC under the highest SCR condition in practice.

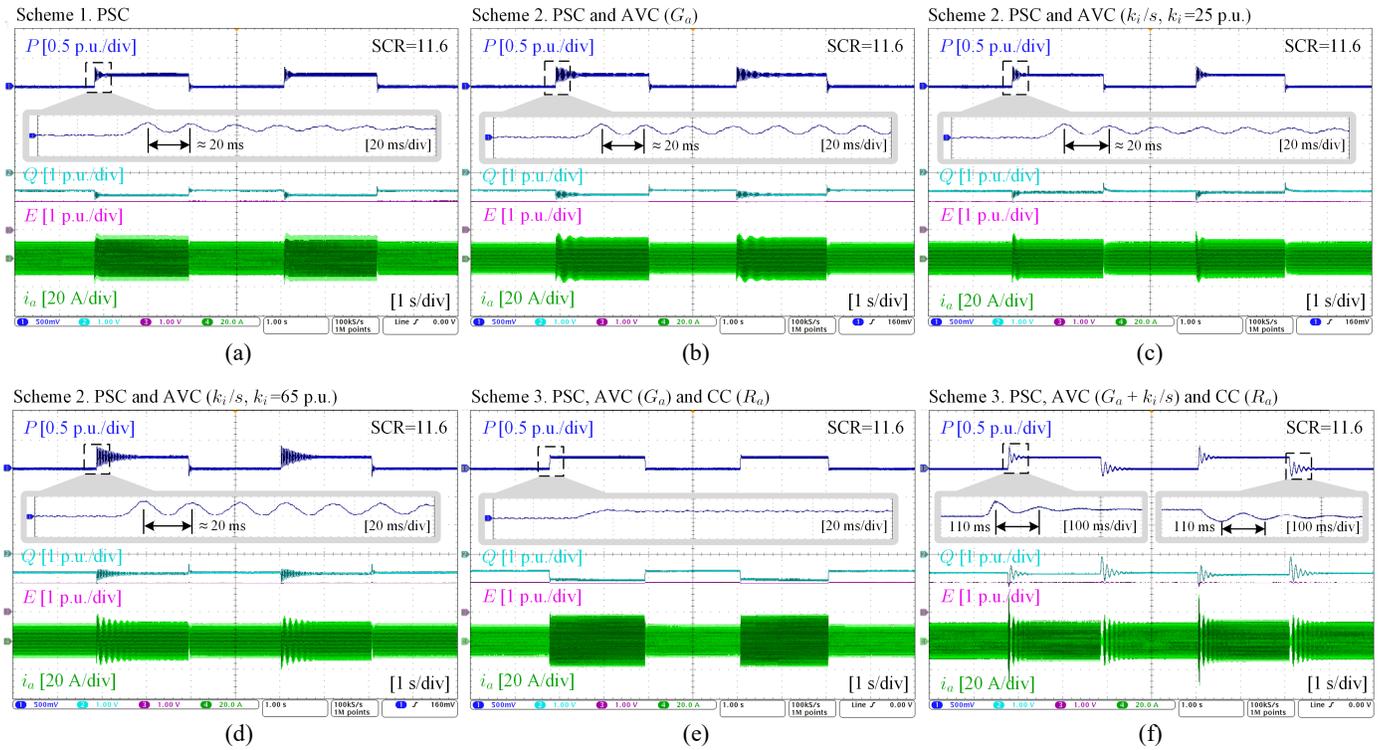

Fig. 12. Comparison of active power responses of 3 control schemes in Table I when SCR=11.6. (a) Scheme 1. (b) Scheme 2 (AVC using P controllers). (c) Scheme 2 (AVC using I controllers and $k_i$=25 p.u.). (d) Scheme 2 (AVC using I controllers and $k_i$=65 p.u.). (e) Scheme 3 (AVC using P controllers). (f) Scheme 3 (AVC using PI controllers).

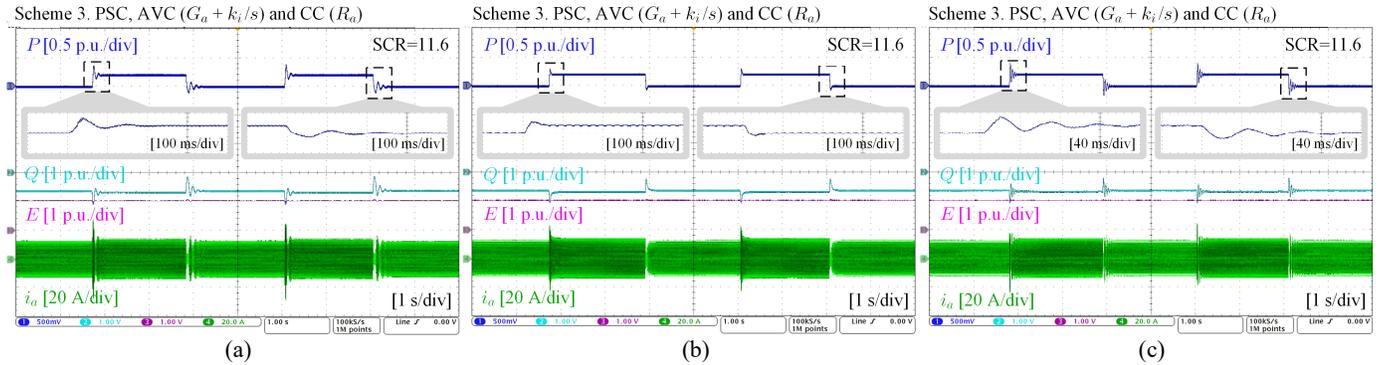

Fig. 13. Comparison of active power responses of scheme 3 in Table I with different control parameters. (a) Reducing $R_a$ to 0.346 p.u. (b) Increasing $G_a$ to 5 p.u. (c) Increasing $k_i$ to 350 p.u.

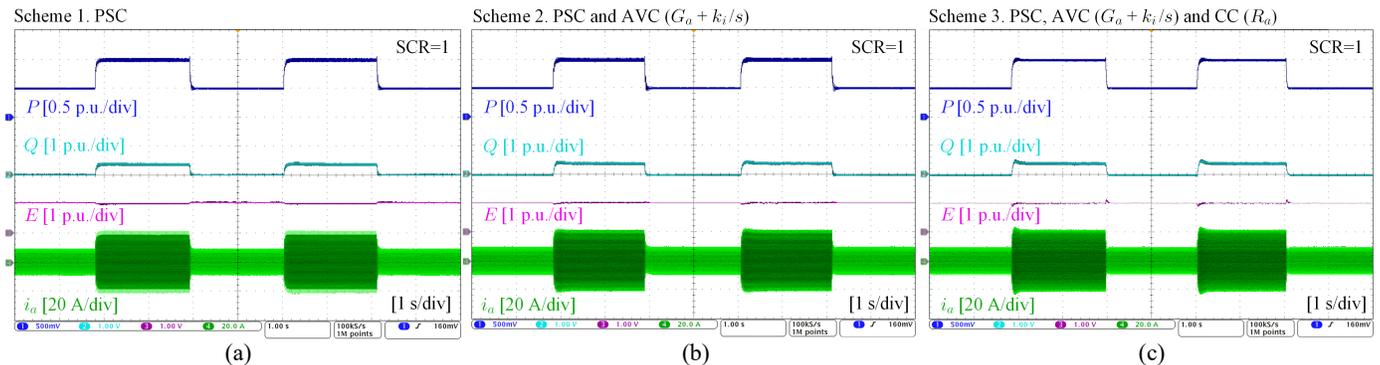

Fig. 14. Comparison of active power responses of 3 schemes in Table I when SCR=1. (a) Scheme 1. (b) Scheme 2. (c) Scheme 3.

## IX. APPENDIX I

Consider a cubic equation

$$s^3+(n_1+m_2)s^2+(n_0+m_1)s+m_0=0 \quad (34)$$

When the following conditions hold, i.e.,

$$n_0 \gg m_1, \ n_0 \gg \varepsilon = (n_1-m_0/n_0+m_2)m_0/n_0 \quad (35)$$

The factorization of polynomial can be approximated to

$$\begin{aligned} &s^3+(n_1+m_2)s^2+(n_0+m_1)s+m_0 \\ &\approx [s^2+(n_1-m_0/n_0+m_2)s+n_0](s+m_0/n_0) \\ &= s^3+(n_1+m_2)s^2+(n_0+\varepsilon)s+m_0 \end{aligned} \quad (36)$$

$n_0 \approx n_0+m_1 \approx n_0+\varepsilon$ is used with the condition (35). Note that the condition (35) can be easily met when

$$n_0 \gg n_1, \ n_0 \gg m_i, \ i=0,\ 1,\ 2 \quad (37)$$

Consequently, the solutions of (36) are given as

$$p_{1,2} \approx \frac{-n_1+\frac{m_0}{n_0}-m_2}{2} \pm j\sqrt{n_0 - \left(\frac{n_1-m_0/n_0+m_2}{2}\right)^2},$$

$$p_3 \approx -m_0/n_0. \quad (38)$$

## IX. APPENDIX II

This section derives the model of GFM control with PSC and AVC. Substituting (11) and (18) into (8) gives

$$\Delta \mathbf{i} = -j\frac{\mathbf{i}_0 s^2+(2\alpha_a \mathbf{i}_0-\boldsymbol{\beta}_a)s+(\alpha_a^2+\omega_1^2)\mathbf{i}_0+\boldsymbol{\eta}_a}{(s+\alpha_a)^2+\omega_1^2}\Delta\theta \quad (39)$$

where $\alpha_a=R/L_{eqa}$, $L_{eqa}=L+G_a L_g$, $\boldsymbol{\beta}_a=(V+G_a \mathbf{E}_0)/L_{eqa}$ and $\boldsymbol{\eta}_a=-(\alpha_a-j\omega_1)\boldsymbol{\beta}_a$. By substituting (11) and (39) into (10), the equivalent PSC plant is given as

$$\Delta P = \kappa \frac{\tau_{2a}s^2+\tau_{1a}s+\tau_{0a}}{(s+\alpha_a)^2+\omega_1^2}\Delta\theta = G_{\theta Pa}(s)\Delta\theta,$$

$$\tau_{2a}=L_g \mathrm{Im}\{\mathbf{i}_0 \boldsymbol{\beta}_a^*\},$$

$$\tau_{1a}=\mathrm{Im}\{\mathbf{E}_0 \boldsymbol{\beta}_a^*\}-\omega_1 L_g \mathrm{Re}\{\mathbf{i}_0 \boldsymbol{\beta}_a^*\}+L_g \mathrm{Im}\{\mathbf{i}_0^* \boldsymbol{\eta}_a\},$$

$$\tau_{0a}=\mathrm{Im}\{\mathbf{E}_0^* \boldsymbol{\eta}_a\}+\omega_1 L_g \mathrm{Re}\{\mathbf{i}_0 \boldsymbol{\eta}_a^*\}. \quad (40)$$

Comparing with (12), the inclusion of AVC dynamics makes the equivalent PSC plant more complex. By substituting $G_{\theta P}(s)$ in Fig. 2 with $G_{\theta Pa}(s)$, the closed-loop transfer function of PSC is given as

$$\Delta P = \frac{[k_P/(\kappa s)]G_{\theta Pa}(s)}{1+[k_P/(\kappa s)]G_{\theta Pa}(s)}\Delta P_{\mathrm{ref}} = G_{\mathrm{PSC}a}(s)\Delta P_{\mathrm{ref}},$$

$$G_{\mathrm{PSC}a}(s) = \frac{k_P(\tau_{2a}s^2+\tau_{1a}s+\tau_{0a})}{s^3+(2\alpha_a+k_P\tau_{2a})s^2+(\alpha_a^2+\omega_1^2+k_P\tau_{1a})s+k_P\tau_{0a}}. \quad (41)$$

$G_{\mathrm{PSC}a}(s)$ has three poles. Define the coefficients $n_{1a}=2\alpha_a$, $n_{0a}=\omega_1^2$, $m_{2a}=k_P\tau_{2a}$, $m_{1a}=\alpha_a^2+k_P\tau_{1a}$ and $m_{0a}=k_P\tau_{0a}$, and substitute them into the denominator of $G_{\mathrm{PSC}a}(s)$. Then, the poles of $G_{\mathrm{PSC}a}(s)$ can be solved by setting the denominator equal to 0, i.e.,

$$s^3+(n_{1a}+m_{2a})s^2+(n_{0a}+m_{1a})s+m_{0a}=0 \quad (42)$$

Using the rules in Appendix I, the closed-form solutions of the three poles can be approximately derived as

$$G_{\mathrm{PSC}a}(s) \approx \frac{k_P(\tau_{2a}s^2+\tau_{1a}s+\tau_{0a})}{(s-p_{1a})(s-p_{2a})(s-p_{3a})}\Delta P_{\mathrm{ref}},$$

$$p_{1,2a}=-\alpha_a+\frac{k_P\tau_{0a}}{2\omega_1^2}-\frac{k_P\tau_{2a}}{2}\pm j\sqrt{D},$$

$$p_{3a}=-k_P\tau_{0a}/\omega_1^2$$

$$D=\omega_1^2-(2\alpha_a-k_P\tau_{0a}/\omega_1^2+k_P\tau_{2a})^2/4. \quad (43)$$

Further, considering that $\alpha_a \ll \omega_1$ and $k_P \ll \omega_1$, the approximations $\boldsymbol{\eta}_a \approx j\omega_1 \boldsymbol{\beta}_a$ and $D \approx \omega_1^2$ can be used. Then, with (40), the final closed-form solutions of the poles are given as

$$p_{1,2a} \approx -\frac{R-\delta}{L+G_a L_g} \pm j\omega_1, \ \delta=k_P(VE_{d0}+G_a|\mathbf{E}_0|^2)/(2\omega_1),$$

$$p_{3a}=-\frac{2\delta+k_P L_g(Vi_{q0}+G_a \mathrm{Im}\{\mathbf{E}_0^* \mathbf{i}_0\})}{L+G_a L_g}. \quad (44)$$

Finally, substituting the coefficients into $m_i$ and $n_i$ ($i=0,\ 1,\ 2$) in gives the solutions (15).

## IX. APPENDIX III

This section derives the model of GFM control with PSC, AVC (P controller), CC and VFF. Substituting (11) and (23) into (8) gives

$$\Delta \mathbf{i} = -j\frac{\mathbf{i}_0 s^2+(\alpha_b \mathbf{i}_0-\boldsymbol{\beta}_b)s+(\omega_1^2+j\omega_1\alpha_b)\mathbf{i}_0-(\alpha_b-j\omega_1)\boldsymbol{\beta}_b}{(s+\alpha_b)^2+\omega_1^2}\Delta\theta \quad (45)$$

where $\alpha_b=(R+R_a)/L_{eqb} \approx R_a/L_{eqb}$, $L_{eqb}=L+(R_a G_a-1)L_g$ and $\boldsymbol{\beta}_b=[V+(R_a G_a-1)\mathbf{E}_0]/L_{eqb}$.

Note that $R_a$ acts as a virtual resistance in series with $L$, and it is much higher than the ESR $R$, thus $R$ can be omitted. In contrast, the AVC controller gain $G_a$ increases the equivalent inductance, as indicated by $L_{eqb}$, while the voltage feedforward has a counteractive effect (the term "-1" in $L_{eqb}$).

By substituting (11) and (45) into (10), the equivalent PSC plant, which includes the AVC and CC dynamics, is given as

$$\Delta P = \kappa \frac{\tau_{2b}s^2+\tau_{1b}s+\tau_{0b}}{(s+\alpha_b)^2+\omega_1^2}\Delta\theta = G_{\theta Pb}(s)\Delta\theta,$$

$$\tau_{2b}=L_g \mathrm{Im}\{\mathbf{i}_0 \boldsymbol{\beta}_b^*\},$$

$$\tau_{1b}=-\alpha_b \mathrm{Im}\{\mathbf{E}_0^* \mathbf{i}_0\}+\mathrm{Im}\{\mathbf{E}_0 \boldsymbol{\beta}_b^*\}$$
$$-\omega_1 L_g \mathrm{Re}\{\mathbf{i}_0 \boldsymbol{\beta}_b^*\}+L_g \mathrm{Re}\{j\mathbf{i}_0 \boldsymbol{\eta}_b^*\}, \quad (46)$$

$$\tau_{0b} = \omega_1 \alpha_b \mathrm{Re}\{\mathbf{E}_0^* \mathbf{i}_0\} + \mathrm{Re}\{j\mathbf{E}_0 \boldsymbol{\eta}_b^*\} + \omega_1 L_g \mathrm{Im}\{j\mathbf{i}_0 \boldsymbol{\eta}_b^*\}$$
$$- (\mathrm{Im}\{\mathbf{E}_0^* \mathbf{i}_0\} + \omega_1 L_g |\mathbf{i}_0|^2) \alpha_b^2$$

where $\boldsymbol{\eta}_b = -(\alpha_b - j\omega_1)\boldsymbol{\beta}_b$. By substituting $G_{\theta P}(s)$ in Fig. 2 with $G_{\theta Pb}(s)$, the closed-loop transfer function is given as

$$\Delta P = \frac{[k_P/(\kappa s)]G_{\theta Pb}(s)}{1 + [k_P/(\kappa s)]G_{\theta Pb}(s)} \Delta P_{\mathrm{ref}} = G_{\mathrm{PSC}b}(s) \Delta P_{\mathrm{ref}}$$
$$= \frac{k_P(\tau_{2b}s^2 + \tau_{1b}s + \tau_{0b})}{s^3 + (2\alpha_b + k_P\tau_{2b})s^2 + (\alpha_b^2 + \omega_1^2 + k_P\tau_{1b})s + k_P\tau_{0b}} \Delta P_{\mathrm{ref}}.$$
(47)

$G_{\mathrm{PSC}b}(s)$ has three poles. Define the coefficients $n_{1b} = 2\alpha_b$, $n_{0b} = \alpha_b^2 + \omega_1^2$, $m_{2b} = k_P\tau_{2b}$, $m_{1b} = k_P\tau_{1b}$ and $m_{0b} = k_P\tau_{0b}$, and substitute them into the denominator of $G_{\mathrm{PSC}b}(s)$. Then, the poles of $G_{\mathrm{PSC}b}(s)$ can be solved by setting the denominator equal to 0, i.e.,

$$s^3 + (n_{1b} + m_{2b})s^2 + (n_{0b} + m_{1b})s + m_{0b} = 0 \quad (48)$$

Using the rules in Appendix I, the closed-form solutions of the three poles can be approximately derived as

$$G_{\mathrm{PSC}b}(s) \approx \frac{k_P(\tau_{2b}s^2 + \tau_{1b}s + \tau_{0b})}{(s - p_{1b})(s - p_{2b})(s - p_{3b})} \Delta P_{\mathrm{ref}}$$
$$p_{1,2b} = -\alpha_b + \frac{k_P\tau_{0b}}{2(\alpha_b^2 + \omega_1^2)} - \frac{k_P\tau_{2b}}{2} \pm j\sqrt{F},$$
$$p_{3b} = -k_P\tau_{0b}/(\alpha_b^2 + \omega_1^2)$$
$$F = \alpha_b^2 + \omega_1^2 - [2\alpha_b - k_P\tau_{0b}/(\alpha_b^2 + \omega_1^2) + k_P\tau_{2b}]^2/4. (49)$$

Since CC gives a high value of virtual resistance $R_a$, the coefficient $\alpha_b \approx R_a/L_{eqb}$ is comparable to $\omega_1$ and cannot be omitted. Then, considering $k_P \ll \omega_1$, one can simply finds that

$$\alpha_b \gg \frac{k_P\tau_{0b}}{2(\alpha_b^2 + \omega_1^2)} - \frac{k_P\tau_{2b}}{2} \quad (50)$$

and $F \approx \omega_1$. As a result, the conjugate poles $p_{1,2b}$ can be further simplified to

$$p_{1,2b} \approx -\alpha_b \pm j\omega_1 = -\frac{R + R_a}{L + (R_a G_a - 1)L_g} \pm j\omega_1 \quad (51)$$

(51) shows that $p_{1,2b}$ are very closed to the poles of equivalent PSC plant in (46).

## IX. APPENDIX IV

This section derives the model of GFM control with PSC, AVC (PI controller), CC and VFF. Substituting (11) and (27) into (8) gives

$$\Delta \mathbf{i} = -j\frac{[s + \alpha_{c0} + j\omega_1(1 + \mu/s)]\mathbf{i}_0 - \boldsymbol{\beta}_c(s)}{s + \alpha_c + j\omega_1(1 + \mu/s)} \Delta\theta \quad (52)$$

where

$$L_{eqc} = L + (R_a G_a - 1)L_g,$$
$$\mu = R_a k_i L_g/L_{eqc},$$
$$\alpha_{c0} = R/L_{eqc} + \mu,$$
$$\alpha_c = (R + R_a)/L_{eqc} + \mu \approx R_a/L_{eqc} + \mu,$$
$$\boldsymbol{\beta}_c(s) = [V + (R_a G_a - 1)\mathbf{E}_0 + R_a k_i \mathbf{E}_0/s]/L_{eqc}. \quad (53)$$

Considering the light-load conditions and substituting $\mathbf{i}_0 \approx 0$ into (52) gives

$$\Delta \mathbf{i} \approx j\frac{\boldsymbol{\beta}_c(s)}{s + \alpha_c + j\omega_1(1 + \mu/s)} \Delta\theta \quad (54)$$

Then focusing on the lower-frequency range (< 10 Hz) where the magnitude of $s$ is normally much lower than $\omega_1\mu/s$ in (54), we can further simplify it by omitting $s$ in the denominator to

$$\Delta \mathbf{i}|_{f<10\,\mathrm{Hz}} \approx j\frac{\boldsymbol{\beta}_c(s)}{\alpha_c + j\omega_1(1 + \mu/s)} \Delta\theta. \quad (55)$$

In addition, with $\mathbf{i}_0 \approx 0$, the steady-state PCC voltage vector becomes

$$\mathbf{E}_0 = V - [(s + j\omega_1)L_f + R]\mathbf{i}_0 \approx V. \quad (56)$$

By substituting (55), (56) and $\mathbf{i}_0 \approx 0$ into (10), the equivalent PSC plant is given as

$$\Delta P = G_{\theta Pc}(s)\Delta\theta \approx \kappa V \Delta i_d$$
$$= \kappa \frac{V}{\alpha_c^2 + \omega_1^2} \frac{\gamma\omega_1 s^2 + (\tau + \mu\gamma)\omega_1 s + \omega_1\mu\tau}{s^2 + \frac{2\omega_1^2\mu}{\alpha_c^2 + \omega_1^2}s + \frac{\omega_1^2\mu^2}{\alpha_c^2 + \omega_1^2}} \Delta\theta \quad (57)$$

where $\gamma = R_a G_a V/L_{eqc}$ and $\tau = R_a k_i V/L_{eqc}$.

By substituting $G_{\theta P}(s)$ in Fig. 2 with $G_{\theta Pc}(s)$, the closed-loop transfer function is given as

$$\Delta P = \frac{[k_P/(\kappa s)]G_{\theta Pc}(s)}{1 + [k_P/(\kappa s)]G_{\theta Pc}(s)} \Delta P_{\mathrm{ref}} = G_{\mathrm{PSC}c}(s) \Delta P_{\mathrm{ref}}$$
$$= \frac{m_{2c}s^2 + n_{0c}s + m_{0c}}{s^3 + (n_{1c} + m_{2c})s^2 + (n_{0c} + m_{1c})s + m_{0c}} \Delta P_{\mathrm{ref}} \quad (58)$$

where

$$m_{2c} = \frac{k_P V\gamma\omega_1}{\alpha_c^2 + \omega_1^2}, \quad m_{1c} = \frac{\omega_1^2\mu^2}{\alpha_c^2 + \omega_1^2}, \quad m_{0c} = \frac{k_P V\omega_1\mu\tau}{\alpha_c^2 + \omega_1^2},$$
$$n_{1c} = \frac{2\omega_1^2\mu}{\alpha_c^2 + \omega_1^2}, \quad n_{0c} = \frac{k_P V(\tau + \mu\gamma)\omega_1}{\alpha_c^2 + \omega_1^2}. \quad (59)$$

Note that $n_{0c}$ is defined as a coefficient in the nominator, which is different from previous schemes. This is to meet the conditions in Appendix I.

Then, the poles of $G_{\mathrm{PSC}c}(s)$ can be solved by setting the denominator equal to 0, i.e.,

$$s^3 + (n_{1c} + m_{2c})s^2 + (n_{0c} + m_{1c})s + m_{0c} = 0 \quad (60)$$

Using the rules in Appendix I with the coefficients in (53) and (59), the closed-form solutions of the three poles can be approximately derived as

$$G_{\text{PSC}c}(s) \approx \frac{m_{2c}s^2 + n_{0c}s + m_{0c}}{(s-p_{1c})(s-p_{2c})(s-p_{3c})} \Delta P_{\text{ref}}$$

$$p_{1,2c} = (\rho_1 - \rho_2)\frac{R_a k_i L_g}{2L_{eqc}} - \frac{k_P G_a \psi}{2}$$
$$\pm j\sqrt{\frac{k_i k_P \psi}{\rho_1} - \left[(\rho_1-\rho_2)\frac{R_a k_i L_g}{2L_{eqc}} - \frac{k_P G_a \psi}{2}\right]^2},$$

$$p_{3c} = -\frac{R_a k_i L_g}{L_f + 2R_a L_g G_a} \quad (61)$$

where

$$\rho_1 = \frac{L_f + R_a L_g G_a}{L_f + 2R_a L_g G_a}, \quad \rho_2 = \frac{2}{\left[\frac{R_a(1+k_i L_g)}{\omega_1 L_{eqc}}\right]^2 + 1},$$

$$\psi = \frac{V^2 R_a/\omega_1 L_{eqc}}{\left[\frac{R_a(1+k_i L_g)}{\omega_1 L_{eqc}}\right]^2 + 1} \quad (62)$$

## IX. Appendix V

It is apparent that the imaginary part of $p_{1,2c}$ in (61) satisfies

$$\sqrt{\frac{k_i k_P \psi}{\rho_1} - \left[(\rho_1-\rho_2)\frac{R_a k_i L_g}{2L_{eqc}} - \frac{k_P G_a \psi}{2}\right]^2} < \sqrt{\frac{k_i k_P \psi}{\rho_1}} \quad (63)$$

When the grid is stiff, i.e., $L_g \approx 0$, the coefficient $\rho_1$ in (62) can be approximated to 1. Then substitute $\psi$ in (62) and $\rho_1 \approx 1$ into (63) gives

$$\sqrt{\frac{k_i k_P \psi}{\rho_1}} \approx \sqrt{k_i k_P \frac{V^2 R_a \omega_1 L_{eqc}}{R_a^2(1+k_i L_g)^2 + \omega_1^2 L_{eqc}^2}} \quad (64)$$

For a typical design of CC gain $R_a \approx 1$ (to give a bandwidth around 500 Hz [23]), it is simple to find $R_a(1+k_i L_g) > \omega_1 L_{eqc}$. Hence, (64) has

$$\sqrt{k_i k_P \frac{V^2 R_a \omega_1 L_{eqc}}{R_a^2(1+k_i L_g)^2 + \omega_1^2 L_{eqc}^2}} < \sqrt{k_i k_P \frac{V^2 R_a \omega_1 L_{eqc}}{2\omega_1^2 L_{eqc}^2}}$$
$$= \sqrt{k_i k_P \frac{V^2 R_a}{2\omega_1 L_{eqc}}} < \sqrt{k_i k_P \frac{V^2 R_a}{2\omega_1 L_f}} \quad (65)$$

In addition, the droop value is often low, e.g., $k_P \leqslant 0.05\omega_1$. The $L$-filter of converter is often around 0.1 p.u., thus $R_a/L_f$ is close to $10\omega_1$. Then, with these conditions, (65) finally has

$$\sqrt{k_i k_P \frac{V^2 R_a}{2\omega_1 L_f}} \leqslant \sqrt{0.25 k_i \omega_1 V^2} \quad (66)$$

## IX. Appendix VI

Substituting $L_{eqc}$ in (53) and (62) into $\rho_2 > 1$ gives

$$\rho_2 > 1 \Rightarrow \frac{R_a(1+k_i L_g)}{\omega_1(L_f + R_a G_a L_g)} < 1$$
$$\Rightarrow (G_a - k_i/\omega_1)\omega_1 L_g > 1 - \omega_1 L_f/R_a. \quad (67)$$

If the parameters are designed to meet $G_a - k_i/\omega_1 > 0$, then (67) has

$$\omega_1 L_g > \frac{1 - \omega_1 L_f/R_a}{G_a - k_i/\omega_1}. \quad (68)$$

With SCR calculated in (1), we can find

$$\text{SCR} < \frac{G_a - k_i/\omega_1}{1 - \omega_1 L_f/R_a}. \quad (69)$$